\documentclass{article}

\usepackage{epsfig}
\usepackage{amsmath}
\usepackage{amssymb}
\usepackage{amsfonts}
\usepackage{mathptmx}
\usepackage{dcolumn}
\usepackage{eucal}
\usepackage{bm}
\usepackage{color}
\usepackage{graphicx}
\usepackage{epstopdf}

\begin{document}

\author{Luca Fasolo$^{1,2}$, Angelo Greco$^{1,2}$, Emanuele Enrico$^{1}$}

\date{%
    \small{$^1$INRiM - Istituto Nazionale di Ricerca Metrologica, Strada delle Cacce 91 10135 Torino, Italy\\%
    $^2$Politecnico di Torino, Corso Duca degli Abruzzi 24, 10129 Torino, Italy\\%
    \vspace{0.5cm}
    \today}
}

\title{Superconducting Josephson-based metamaterials for quantum-limited parametric amplification: a review}

\maketitle

\begin{abstract}
In the last few years, several groups have proposed and developed their own platforms demonstrating quantum-limited linear parametric amplification, with evident applications in quantum information and computation, electrical and optical metrology, radio astronomy and basic physics concerning axion detection.
Here we propose a short review on the physics behind parametric amplification via metamaterials composed by coplanar wave-guides embedding several Josephson junctions. We present and compare different schemes that exploit the nonlinearity of the Josephson current-phase relation to  mix the so-called signal, idler and pump tones. The chapter then presents
and compares three different theoretical models, developed in the last few years, to predict the dynamics of these nonlinear systems in the particular case of a 4-Wave Mixing process and under the degenerate undepleted pump assumption.
We will demonstrate that, under the same assumption, all the results are comparable in terms of amplification of the output fields. 
\end{abstract}

\section{Introduction}

In the last decade microwave quantum electronics received a substantial boost by the advancements in superconducting circuits and dilution refrigerators technologies. These platforms allow experiments to be easily carried out in the mK regime, where the detection and manipulation of signals in the range $3-12$ GHz reaches energy sensitivities comparable to a single photon \cite{Schoelkopf2013}.

Solid state microwave quantum electronics is founded on a building block that has no analogous in quantum optics: the Josephson junction \cite{Josephson1962}. This, in fact, is a unique nondissipative and nonlinear component that represents the key element of a large series of quantum experiments.

Furthermore, microwave quantum electronics allows the exploration of the so-called ultrastrong coupling regime \cite{Niemczyk2010}, hard to be reached in quantum optics, and it is worth mentioning that nonlinear resonator can be exploited to access relativistic quantum effects and quantum vacuum effects. To give an example, the Lamb shift \cite{Lamb1947} effect has been observed in superconducting artificial atom \cite{Fragner2008} while the dynamical Casimir effect \cite{Moore1970,Fulling1976} has been promoted by properly engineered superconducting waveguide \cite{Wilson2011}.

From the very beginning, superconducting electronics has been pushed by the strong interest coming from the quantum computation and information community. However, it has been only recently shown that a new concept of 1D metamaterial with embedded several Josephson junctions enables strong photon-photon on-chip interactions \cite{Ketchen1991}, allowing experimentalists to engineer dispersion relations that drive the waves travelling along artificial waveguides \cite{O'Brien2014, White2015}.
These concepts and technologies allow the control and tunability of the wave mixing process. As an example, a weak signal travelling in a metamaterial can interact with a strong pump tone at a different frequency, activating the so called parametric amplification \cite{Cullen1959}.
The class of devices where these phenomena are promoted is commonly known as Travelling Wave Josephson Parametric Amplifiers (TWJPA) and represents the solid state analogous to optical $\chi^n$ nonlinear crystals \cite{Sweeny1985}.

It has been shown that TWJPAs can act as quantum parametric amplifiers by reaching the so-called quantum limit \cite{Macklin2015}. With the purpose of a comparison to the stat-of-the-art commercially available low-noise amplifiers, these latter can operate at $\omega/2\pi =$ 4 GHz adding $k_BT_n/\hbar\omega\approx10$ noise photons having a noise temperature of $T_n=2$ K, while Josephson-based amplifiers can reduce this added noise up to $1/2$ photon, or even 0, depending on its working configuration.

The capability to beat the quantum limit is related to the so-called phase-sensitive amplification process, where the metamaterial can operate in degenerate mode (degenerate parametric amplifier, DPA), acting on two waves (signal and idler) at the same frequency ($\omega_s=\omega_i$) by amplifying and de-amplifying their position and momentum quadratures respectively. In this view, DPA enables the preparation of squeezed states in the microwave regime.
Even in the non-degenerate mode (non-degenerate parametric amplifier, NDPA, i.e. $\omega_s\neq\omega_i$) the phase-preserving nature of the quantum parametric amplification results in the entanglement condition among the signal and idler generated photons, composing a two-mode squeezed state \cite{Walls2008}. It's worth mentioning how, such a quantum state, is an example of Einstein-Podolsky-Rosen state \cite{EPR1935}, where correlations between signal and idler are stronger than that allowed by classical theory \cite{Reid1988}.

It should be evident how superconducting electronics not only has demonstrated to be an ideal platform for microwaves quantum parametric amplification, but it is pushing forward the research field focused on the generation of nonclassical radiation with attractive potential applications in metrology and quantum information processing.

\section{Historical evolution of the Travelling Wave  Parametric Amplifiers}
\label{sec:history}
\begin{figure}[h!]
    \begin{centering}
    \includegraphics[width=1\textwidth]{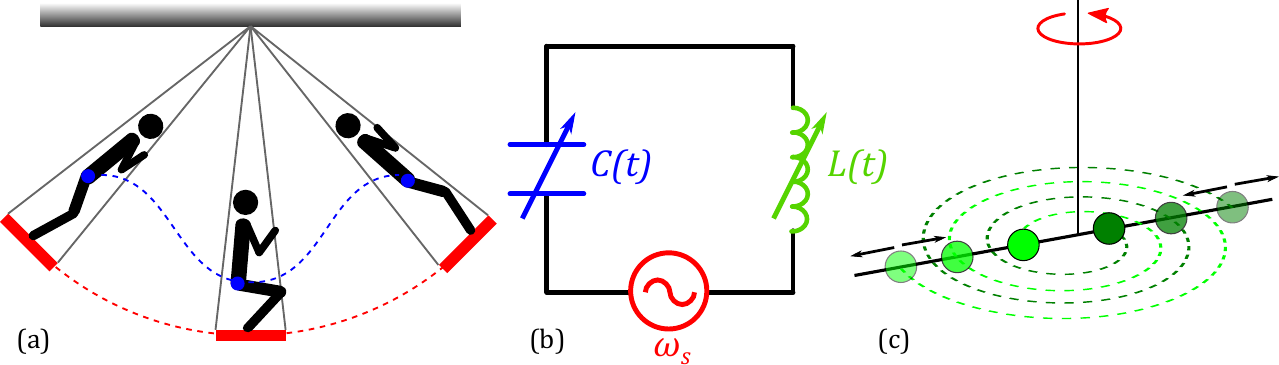}
\par\end{centering}
    \caption{(a) Sketch of a swing process. An oscillating system at a frequency $\omega_s$ is excited by parametric amplification via periodical changes of the center of mass position at a frequency $\omega_p=2\omega_s$. (b) LC circuit with variable (nonlinear) C and L components. The case in which the capacitance C is periodically changed in time is the circuit analogous to the mechanical system represented in (a) while the case having an oscillating inductance L mimics the condition sketched in (c), consisting in a torque pendulum with variable inertia momentum \cite{Butikov2004}.}
    \label{fig:swing}
\end{figure}

The theory of a new concept of microwave amplifier was developed by A. L. Cullen \cite{Cullen1959} in 1959. In his paper Cullen showed a novel mechanism of periodic transfer of power between a pump tone and a signal traveling in a transmission line composed of a voltage dependent capacitance per unit length.
A non-linear component of an RLC circuit can change periodically the resonance frequency of the whole system, leading to a novel way of making broadband amplification, the so called Parametric Amplification. In Figure \ref{fig:swing} we report two toy models for parametric amplification in mechanical systems with their electrical counterparts. 

One of the first realizations of Cullen's idea was made by R. Mavaddat \textit{et al.} in 1962 \cite{Mavaddat1962}. The signal line was basically a low pass filter in which the shunt elements were similar varactor diodes. There the nonlinearity was given by the specific capacitance-voltage relation of the varactor diodes, which is highly non-linear for relative small voltage values. In this pioneering experiment a gain of 10 dB and a bandwidth of 3 MHz were shown.\\
After the theorizing and the subsequent discovery of the Josephson effect \cite{Josephson1962}, it was understood that an easy way to embed a non-linear component into a transmission line and simultaneously reduce losses, was to build a non-linear inductance made of superconducting material, exploiting a Josephson junction as a source of non-linearity following the vanguard idea by M. Sweeny and R. Mahler \cite{Sweeny1985}. There the parametric amplifier was modeled by a first-order small-signal theory with the same approach adopted to predict the behaviour of GaAsFET transmission line amplifiers.
The proposed design consisted of a superconducting thin-film niobium transmission line, composed by a coplanar waveguide integrating a large number of Josephson junctions.\\
The first realization of a Travelling Wave Parametric Amplifier (TWJPA) embedding a series of Josephson junctions was possible due to the PARTS process developed at IBM \cite{Ketchen1991}. Exploiting niobium/aluminum technology, Yurke \textit{et al.} \cite{Yurke1996} reported the construction and characterization of a coplanar waveguide in which the central trace was composed by an array of 1000 Josephson junctions. The experiment was there performed in reflection mode, by terminating one end of the device with a short, leading to a relative high gain of 16 dB but a narrow bandwidth of 125 MHz and a noise temperature of 0.5 $\pm$ 0.1 K. The mismatch between the theoretical model and the experimental data has resulted in the understanding of a lack of a complete description of the physics behind this device when working in a small-signal regime.
The study of the collective behaviour of groups of Josephson junctions forming a transmission line has been an active field of study of several theoretical works \cite{Zant1994,Caputo1997}. Subsequently, the use of numerical analysis \cite{Mohebbi2009} helped clarifying how wave propagation acts inside this kind of transmission line, giving information on cutoff propagation, dispersive behavior and shock-wave formation. An analytical model of a Josephson travelling wave amplifier of greater complexity was developed by Yaakobi \textit{et al.} \cite{Yaakobi2013}. There a transmission line made of a series of capacitively shunted Josephson junctions was considered.\\ 
One of the main limitation concerning the maximum achievable gain, common to all the TWJPAs concepts, is represented by the phase mismatch between the different tones into the line. In particular, it has to be noticed that even though the incoming waves can be in phase, photon-photon interactions between different tones (cross-phase modulation) or the same tone (self-phase modulation) lead to a modification of the phase of the travelling tones themselves.
Indeed, quantum mechanically speaking, the power transport between the pump and the signal waves takes place through a photon energy conversion between the pump and the signal. This means that, for an efficient energy exchange, conservation of both energy and momentum needs to take place. The latter condition is the corpuscular analogous to the phase matching requirement between the different electromagnetic waves.
An engineering solution to overcome this problem is represented by the so-called Resonant Phase Matching (RPM) \cite{O'Brien2014}. O'Brien \textit{et al.} analyzed this method theoretically on a simple transmission line made of a series of Josephson junctions capacitively shunted to ground operating in the so-called 4-Wave Mixing (4WM) regime. In their model they shunted the transmission line with several LC resonators with a resonance frequency slightly above the pump tone. Doing this they were able to show the arise of a stop band in the dispersion relation, that is able to re-phase the pump with the signal tones  by changing the pump wave vector, favoring the wave mixing.\\
O'Brien's design was realized not long after \cite{White2015} using Al technology. In their design the unit cell of the transmission line was composed by three single non-linear Josephson cells, the shunt capacitor was made using low-loss amorphous silicon dielectric and a resonator was placed after each group of 17 unit cells. The device showed a maximum gain of 12 dB over a 4 GHz bandwidth centered on $\approx$ 5 GHz. Moreover, the authors explain that variations of 2 - 3 dB in the gain most likely come from imperfect impedance matching between sections and at the level of the bond pads.\\
A similar design was adopted by Macklin \textit{et al.} \cite{Macklin2015} to prove experimentally the capability of a TWJPA combined with the RPM technique to be used as a reliable tool for qubits readout. In this paper the TWJPA, based on Nb technology and a different RPM periodicity, was first characterized, showing a gain of 20 dB over a 3 GHz bandwidth. Moreover, the quantum efficiency of the amplifier was tested when coupled with a 3D Transmon qubit, leading to an efficiency value of $0.49 \pm 0.01$. A key point of this experiment was the proof that a single TWJPA could be able to perform the readout of more than 20 qubits thanks to its high dynamic range and multiplexing capabilities.
RPM has shown remarkable capabilities and is a promising technique to overcome phase mismatch. It can be implemented in multiple ways \cite{Tan2017}, by the way, it has to be noticed that this method requires an increase of design complexity, lower tolerances on the constructing parameters and longer propagation lengths (2 cm - 1 m).\\
Another option to solve the mismatch problem was suggested by Bell and Samlov \cite{Bell2015}, who proposed a self-phase matching transmission line embedding a series of asymmetric superconducting quantum interference devices (SQUIDs). The remarkable feature of this design is that it does not need any resonant circuit to achieve phase matching. This TWJPA is indeed able to tune the non-linearity of its SQUIDs just through the use of an external magnetic field. Zhang \textit{et al.} realized this design \cite{Zhang2017} proving the widely tunability on positive and negative values of the Kerr nonlinearity by a magnetic flux and its capability to assist phase matching in the 4WM process. 
The 4WM process is intrinsically affected by phase mismatch because it takes origin from a cubic (Kerr-like) nonlinearity of the current-phase relation of the SQUIDs composing the TWJPA, getting unwanted effects from Self-Phase and Cross-Phase modulations.\\
A. B. Zorin showed \cite{Zorin2016} that embedding a chain of rf-SQUIDs into a coplanar waveguide it is possible to tune both the second and third order nonlinearities of their phase-current relation. This is a totally novel approach to the TWJPA since, the possibility to use a quadratic term as a source of nonlinearity, allows to work in the 3-Wave Mixing regime (3WM) as theorized by Cullen 57 years before.
It's well known that 3WM has several advantages compared to 4WM. Firstly, it allows to operate with a minimal phase mismatch. Secondly, it requires a lesser pump power to achieve the same amplification per unit length. Eventually, it separates signal and idler from pump tones, easing the engineering of the experimental setup by removing the requirement of heavy filtering in the middle of the amplification band. A proof of principle based on the Zorin's layout \cite{Zorin2017}, showed a gain reaching 11 dB over a 3 GHz bandwidth.\\
A step forward in controlling the metamaterial nonlinearities was attempted by Miano \textit{et al.} \cite{Miano2019} achieving an independent tune of both second and third order terms in the current-phase relation by adjusting the bias current in some inductive circuits surrounding the transmission line. This technology takes the name of Symmetric Travelling Wave Parametric Amplifier (STWPA), its peculiarity arising from the symmetric arrangement of the rf-SQUIDs that compose the transmission line. This device concept represents the state-of-the-art in the field, allowing the exploration a wide portion of the control parameters space, leading to a maximum estimated gain of 17 dB and a 4 GHz bandwidth.

\section{Theoretical models for a 4WM process in a TWJPA}
\label{sec:models}
\begin{figure}[h!]
 \begin{centering}
    \includegraphics[width=1\textwidth]{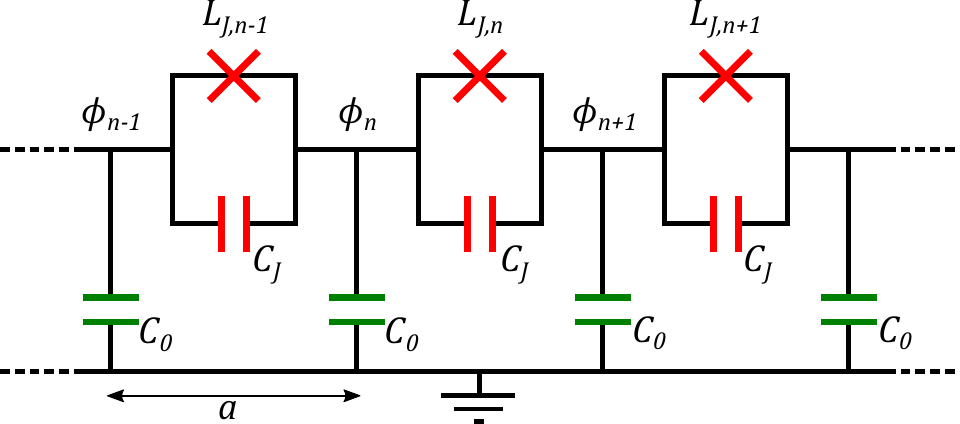}
\par\end{centering}
    \caption{Electrical equivalent representation of a repetition of Josephson junctions embedded in a transmission line. The junctions are modelled as an LC resonant circuit. The length of the unitary cell length is reppresente by $a$.}
    \label{fig:Elementary_cell}
\end{figure}

In the last decade different theoretical models have been developed to predict the behaviour of an electric transmission line containing an array of Josephson junctions, employed as nonlinear elements. In this section we will focus on those models developed to predict the behaviour of a TWJPA in the particular case of a 4WM process, under undepleted degenerate pump approximation (i.e. assuming that the power holded by the pump wave is at first approximation constant and larger than the one owned by the signal and the idler). We will firstly focus on the classical theory proposed by Yaakobi \textit{et al.} in 2013 \cite{Yaakobi2013} and O'Brien \textit{et al.} in 2014 \cite{O'Brien2014}, in which the behaviour of the transmission line is derived imposing the current conservation in the system. This starting assumption leads to the definition of a partial differential nonlinear equation that can be turned into a system of coupled mode equations, providing the expression of the amplitude of the pump, signal and idler tones along the transmission line. 
Subsequently, we will discuss two different quantum approaches for the description of the parametric amplifier dynamics. The first one, proposed by Grimsmo and Blais in 2017 \cite{Grimsmo2017}, exploits an Hamiltonian based on continuous-mode operators to derive, in an interaction picture frame, a device's output field. The second one instead, proposed by van der Reep in 2019 \cite{vanderReep2019}, derives a system of coupled mode equations for the creation and annihilation quantum operators starting from an Hamiltonian based on discrete-mode operators.\\
The theories presented in this chapter will be based on a series of simplifying assumptions, whose experimental realisation could be difficult to be obtained. For instance, in a real device, the undepleted pump approximation is hardly respected along the entire extension of the device because, along the line, the pump tone transfers a non-negligible amount of energy to the signal and idler one. The depletion effects, resulting in a reduction of the gain and of the dynamics-range of the amplifier, have been studied both in a classical and quantum frame \cite{Zorin2016, Roy2018}.\\
In all of these models, a lossless electrical circuit composed by the repetition of an elementary cell, whose structure is shown in Figure \ref{fig:Elementary_cell}, is taken into account. 
In order to standardise the notations, we assume that the Josephson junctions embedded in the transmission line are identical (i.e. they have the same critical current $I_c$) and that the current flowing through the \textit{n}-th junction can be expressed through the nonlinear relation 
\begin{equation}
    \label{eq:Current_Phase_Relation}
    I_{J,n}=I_c\sin{\big(\varphi_n\big)}
\end{equation}
where $\varphi_n(t)$ is the phase-difference across the junction of the macroscopic wave functions of the two superconductive electrodes. The relation between $\varphi_n$ and the voltage-difference across the junction is given by the Faraday's induction law
\begin{equation}
\label{eq:Voltage_Phase_Relation}
    \Delta V_n=V_{n+1}-V_n=\frac{\Phi_0}{2\pi}\;\frac{d\varphi_n}{dt}=\frac{\Phi_0}{2\pi}\;\frac{d}{dt}\Big[\phi_{n+1}-\phi_{n}\Big]
\end{equation}
where $\Phi_0=h/(2e)$ is the magnetic quantum flux (with \textit{h} the Planck constant and \textit{e} the elementary charge) whereas $\phi_n(t)$ is the absolute phase in the \textit{n}-th node of the circuit. The phase at the \textit{n}-th node ($\phi_n$) can be converted into a flux at the \textit{n}-th node, and vice versa, through the relation $\Phi_n=(\Phi_0/2\pi)\phi_n$.\\
Furthermore, we define $C_{J}$ the capacitance associated to the \textit{n}-th Josephson junction and $L_{J,n}$ its inductance, defined as
\begin{equation}
    L_{J,n}=\frac{\Delta V_n}{dI_{J,n}/dt}=\frac{\Phi_0}{2\pi}\;\frac{1}{I_c\cos{(\varphi_n)}}=\frac{L_{J_0}}{\cos(\varphi_n)}
\end{equation}
where $L_{J_0}=\Phi_0/(2\pi I_c)$ is the inductance of the Josephson junction for a phase difference $\varphi_n=0$.\\
The energy stored in the \textit{n}-th Josephson junction can be expressed, using the definitions given in (Eq. (\ref{eq:Current_Phase_Relation})) and (Eq. (\ref{eq:Voltage_Phase_Relation})) as
\begin{equation}
\label{eq:Energy_Stored_JJ}
    U_{J,n}=\displaystyle\int^t_{t_0} VI\;dt'=\displaystyle\int^t_{t_0}\frac{\Phi_0}{2\pi}\frac{d\varphi_n(t')}{dt'}\;I_c\sin{(\varphi_n)}\;dt'=I_c \frac{\Phi_0}{2\pi}\big[1-cos{\big(\varphi_n(t)\big)}\big],
\end{equation}
under the assumption that $\varphi_n(t_0)=0$, and approximated through a first-order power expansion as
\begin{equation}
\label{eq:Josephson_Energy}
    U_{J,n}=I_c \frac{\Phi_0}{2\pi}\big[1-cos{\big(\varphi_n\big)}\big]
    =\frac{1}{2L_{J_0}}\Delta\Phi_n^2-\frac{1}{24L_{J_0}}\bigg(\frac{2\pi}{\Phi_0}\bigg)^2\Delta\Phi_n^4+O\big(\Delta\Phi_n^6\big)
\end{equation}
where $\Delta\Phi_n=\Phi_{n+1}-\Phi_n$.\\
Finally, we assume identical the coupling capacitances $C_0$ between the transmission line and ground.

\subsection{The classical theoretical model}
In this subsection we will present the main steps for the derivation of the classical model presented in \cite{Yaakobi2013,O'Brien2014}. Under proper assumption, this model allows to determine analytically the amplitude of the signal's and idler's waves along the transmission line.\\
Expressing the current flowing through each branch of the circuit presented in Fig. \ref{fig:Elementary_cell} in terms of absolute phases $\phi_n$ and imposing the current conservation in the \textit{n}-th node (i.e. $I_{J,n-1}+I_{C_J,n-1}=I_{J,n}+I_{C_J,n}+I_{C_0,n}$), a differential equation for the absolute phase, in the case of a weak nonlinearity, can be obtained:
\begin{align}
    -C_0\;\frac{d^2}{dt^2}\big[\phi_n\big]&=-C_J\;\frac{d^2}{dt^2}\big[\phi_{n+1}+\phi_{n-1}-2\phi_n\big]-\frac{1}{L_{J_0}}\big[\phi_{n+1}+\phi_{n-1}-2\phi_n\big]+\nonumber\\
    &+\frac{\Phi_0}{2\pi}\frac{1}{6I^2_c\; L_{J_0}^3}\big[\big(\phi_{n+1}-\phi_n\big)^3-\big(\phi_n-\phi_{n-1}\big)^3\big]
\end{align}
where the last term derives from the first-order approximation of the nonlinear behaviour of the Josephson's inductance.\newline
Assuming the length \textit{a} of the elementary cell much smaller than the wave lengths of the propagating waves $\lambda$ (i.e. $a/\lambda \ll 1$), the discrete index \textit{n} can be replaced by a continuous position \textit{x} along the line (i.e. $\phi_n(t) \to \phi(x,t)$) and the phase differences can be expressed, at the second order approximation, as:
\begin{align}
    \phi_{n+1}-\phi_{n}&\approx a\frac{\partial\phi}{\partial x}+\frac{1}{2}\;a^2\frac{\partial^2\phi}{\partial x^2}\\
    \phi_{n}-\phi_{n-1}&\approx a\frac{\partial\phi}{\partial x}-\frac{1}{2}\;a^2\frac{\partial^2\phi}{\partial x^2}
\end{align}
In this way, it is possible to define a nonlinear differential equation for the continuous absolute phase $\phi(x,t)$:
\begin{equation}
\label{eq:Absolute_Phase_Equation}
   C_0\;\frac{\partial^2\phi}{\partial t^2}-\frac{a^2}{L_{J_0}}\frac{\partial^2\phi}{\partial x^2}-C_J\;a^2\;\frac{\partial^4\phi}{\partial x^2 \partial t^2}=-\frac{a^4}{2\;I^2_c\;L^3_{J_0}}\frac{\partial^2\phi}{\partial x^2}\Bigg(\frac{\partial\phi}{\partial x}\Bigg)^2
\end{equation}
In the case of a weakly nonlinear medium, the dispersion law can be derived from (Eq. (\ref{eq:Absolute_Phase_Equation})) considering the left-hand side being equal to zero and imposing a plane-wave solution $\phi(x,t)\propto e^{i\;(kx-\omega t)}$:
\begin{equation}
\label{eq:Dispersion_Law}
    k(\omega)=\frac{\omega\;\sqrt{L_{J_0}C_0}}{a\sqrt{1-L_{J_0}C_J\omega^2}}
\end{equation}
The solutions of (Eq. (\ref{eq:Absolute_Phase_Equation})) can be expressed, as shown by O'Brien \textit{et al.} \cite{O'Brien2014}, in the form of a superposition of three waves (pump, signal and idler) whose amplitudes are complex functions of the position along the line:
\begin{equation}
\label{eq:ClassicalPhases}
    \phi(x,t)=\displaystyle\sum_{n=p,s,i}Re\Big[A_n(x)\;e^{i\;(k_nx-\omega_n t)}\Big]=\frac{1}{2}\displaystyle\sum_{n=s,i,p}\;\Big[A_n(x)\;e^{i\;(k_n x-\omega_n t)}+c.c\Big]
\end{equation}
The case of a 4WM process with a degenerate pump can be taken into account imposing the frequency matching condition $2\omega_p=\omega_s+\omega_i$. Replacing this particular solution in (Eq. (\ref{eq:Absolute_Phase_Equation})) and assuming that, along the line, the amplitudes are slowly varying (i.e. $|\partial^2 A_n/\partial x^2| \ll k_n|\partial A_n/\partial x| \ll k_n^2 |A_n|$) and that $|A_s|^2$ and $|A_i|^2$ are negligible (i.e. $|A_{s,i}| \ll |A_p|$, strong pump approximation), we obtain a system of three coupled differential equations for the amplitudes $A_n(x)$ that describe the energy exchange between the three waves along the line:
\begin{align}
\label{eq:CME_1}
    \frac{\partial A_p}{\partial x}&=i\vartheta_p|A_p|^2A_p+2iX_pA^*_pA_sA_i\;e^{i\Delta k\;x}\\
\label{eq:CME_2}    
    \frac{\partial A_{s(i)}}{\partial x}& =i\vartheta_{s(i)}|A_p|^2A_{s(i)}+iX_{s(i)}A^2_pA_{i(s)}\;e^{i\Delta k\;x}
\end{align}
where $\Delta k=2k_p-k_s-k_i$ is the chromatic dispersion. The term $\vartheta_{p}$ is responsible for the self-phase modulation of the pump tone while $\vartheta_{s(i)}$ are responsible for the cross-phase modulation between the pump tone and the signal or idler respectively. These terms can be expressed as
\begin{equation}
    \vartheta_{p}=\frac{a^4k_p^5}{16C_0I^2_cL^3_{J_0}\omega^2_p}\hspace{0.5cm}\text{and}\hspace{0.5cm}\vartheta_{s(i)}=\frac{a^4k^2_pk^3_{s(i)}}{8C_0I^2_cL^3_{J_0}\omega^2_{s(i)}}
\end{equation}
while the coupling constants $X_n$, depending on the circuit parameters, are defined as
\begin{equation}
    X_p=\frac{a^4k^2_pk_sk_i(k_p-\Delta k)}{16C_0I^2_cL^3_{J_0}\omega^2_p}\hspace{0.5cm}\text{and}\hspace{0.5cm}X_{s(i)}=\frac{a^4k^2_pk_sk_i(k_{s(i)}+\Delta k)}{16C_0I^2_cL^3_{J_0}\omega^2_{s(i)}}
\end{equation}
Expressing the complex amplitudes $A_n(x)$ in a the co-rotating frame
\begin{equation}
    A_n(x)=A_{n_0}e^{\;i\vartheta_n |A_{p_0}|^2x}
\end{equation}
it can be demonstrated that, working under the undepleted pump approximation $|A_p(x)|=A_{p_0}\gg|A_{s(i)}(x)|$,  the amplitude of the signal can be expressed as 
\begin{equation}
\label{eq:Final_1}
    A_s(x)=\Bigg[A_{s_0}\Bigg(\cosh{(g_1x)-\frac{i\;\Psi_1}{2g_1}\sinh{(g_1x)}}\Bigg)+i\frac{X_{s(i)}|A_{p_0}|^2}{g_1}A^{*}_{s(i)_0}\sinh{(g_1x)}\Bigg]e^{\;i\frac{\Psi_1}{2}x}
\end{equation}
where $\Psi_1=\Delta k+(2\vartheta_p-\vartheta_s-\vartheta_i)|A_{p_0}|^2=\Delta k+\vartheta |A_{p_0}|^2$ is the total phase mismatch and $g_1$ is the exponential complex gain factor, defined as
\begin{equation}
    g_1=\sqrt{X_sX^*_i|A_{p_0}|^4-\bigg(\frac{\Psi_1}{2}\bigg)^2}
\end{equation}
The total gain of an amplifier, composed by the repetition of $N$ elementary cells, can then be expressed as $G_s(aN)=|A_s(aN)/A_{s_0}|^2$.


\subsection{Quantum Hamiltonian model based on continuous-mode operators}

A standard method to treat quantum superconducting circuits is represented by the lumped element approach \cite{Vool2017}. In this latter the Hamiltonian of the quantum circuit is straightforwardly derived from its classical counterpart by promoting fields to operators and properly imposing commutating relations. In this view, one can proceed by deriving the Lagrangian of a TWJPA composed by the repetition of N unitary cells, under first nonlinear order approximation, as
\begin{align}
    \mathcal{L}&=\displaystyle\sum^{N-1}_{n=0}\bigg[\frac{C_0}{2}\bigg(\frac{\partial\Phi_n}{\partial t}\bigg)^2+\frac{C_J}{2}\bigg(\frac{\partial\Delta\Phi_n}{\partial t}\bigg)^2-E_{J_0}\bigg(1-\cos{\Big(\frac{2\pi}{\Phi_0}\Delta\Phi_n\Big)}\bigg)\bigg]\approx\\
    & \approx \displaystyle\sum^{N-1}_{n=0}\bigg[\frac{C_0}{2}\bigg(\frac{\partial\Phi_n}{\partial t}\bigg)^2+\frac{C_J}{2}\bigg(\frac{\partial\Delta\Phi_n}{\partial t}\bigg)^2-\frac{1}{2L_{J_0}}\Delta\Phi^2_n-\frac{1}{24L_{J_0}}\bigg(\frac{2\pi}{\Phi_0}\bigg)^2\Delta\Phi^4_n\bigg]
\end{align}
where $E_{J_0}=I_c \Phi_0/2\pi=I_c L_{J_0}$. Under the assumption that $a/\lambda\ll 1$ it is possible, as performed in the previous subsection, to replace the discrete index $n$ with a continuous position $x$ along the line (i.e. $\Phi_n(t)\to\Phi(x,t)$) and approximate, at the first order, $\Delta\Phi_n\to a\;\partial\Phi(x,t)/\partial x$. Furthermore, extending the system via two lossless semi-infinite transmission lines (characterized by a constant distributed capacitance $c_0$ and a constant distributed inductance $l_0$) the Lagrangian can be expressed through a space integral extending from $x=-\infty$ to $x=+\infty$ as
\begin{equation}
\label{eq:Lagrangian}
    \mathcal{L}\bigg[\Phi,\frac{\partial\Phi}{\partial t}\bigg]=\frac{1}{2}\displaystyle\int^{\infty}_{-\infty}\bigg[c(x)\bigg(\frac{\partial\Phi}{\partial t}\bigg)^2+\frac{1}{\omega^2_{J}(x) l(x)}\bigg(\frac{\partial^2\Phi}{\partial x\partial t}\bigg)-\frac{1}{l(x)}\bigg(\frac{\partial\Phi}{\partial x}\bigg)^2+\gamma(x)\bigg(\frac{\partial\Phi}{\partial x}\bigg)^4\bigg]dx
\end{equation}
where $c(x)$ and $l(x)$ are the distributed capacitance and inductance of the system, defined as
\begin{equation}
c(x)=
    \begin{cases}
    c_0     & x<0\\
    C_J/a   & 0<x<z\\
    c_0     &  x>z
    \end{cases}
    \hspace{0.5cm}\text{and}\hspace{0.5cm}
    l(x)=
    \begin{cases}
    l_0     & x<0\\
    L_{J_0}/a & 0<x<z\\
    l_0     & x> z
    \end{cases}
\end{equation}
\begin{equation}
\omega_J(x)=
    \begin{cases}
    \infty     & x<0\\
    1/\sqrt{L_{J_0} C_J}   & 0<x<z\\
    \infty     &  x>z
    \end{cases}
    \hspace{0.5cm}\text{and}\hspace{0.5cm}
    \gamma (x)=
    \begin{cases}
    0     & x<0\\
    (a^3 E_{J_0}/12)(2\pi/\Phi_0)^4 & 0<x<z\\
    0     & x> z
    \end{cases}
\end{equation}
where $z$ is the length of the TWJPA, $\omega_{J}(x)$ is the junction's plasma frequency and $\gamma(x)$ is the term deriving from the nonlinearity of the junctions.\\
From (Eq. (\ref{eq:Lagrangian})) one can easily derive the Euler-Lagrange equation whose form, for $0<x<z$, is equal to (Eq. (\ref{eq:Absolute_Phase_Equation})) giving the same dispersion relation (Eq. (\ref{eq:Dispersion_Law})) under analogous assumptions. Instead, outside the nonlinear region, the wavevector turns out to be $k_\omega(x)=\sqrt{c_0l_0}\omega^2$.\\
The Hamiltonian of the system can be derived from the Lagrangian taking into account $\pi(x,t)=\delta \mathcal{L}/\delta [\partial \Phi/\partial t]$, the canonical momentum  of the flux $\Phi(x,t)$:
\begin{align}
    H&[\Phi,\pi]=\displaystyle\int^{\infty}_{-\infty}\bigg[\pi\frac{\partial\Phi}{\partial t}\bigg]dx-\mathcal{L}=\nonumber\\
    &=\frac{1}{2}\displaystyle\int^{\infty}_{-\infty}\bigg[c(x)\bigg(\frac{\partial\Phi}{\partial \Phi t}\bigg)^2+\frac{1}{l(x)}\bigg(\frac{\partial\Phi}{\partial x}\bigg)^2+\frac{1}{\omega^2_p(x) l(x)}\bigg(\frac{\partial^2\Phi}{\partial x \partial t}
    \bigg)^2\bigg]dx-\frac{\gamma}{2}\displaystyle\int^{z}_{0}\bigg[\bigg(\frac{\partial\Phi}{\partial x}\bigg)^4\bigg]dx=\nonumber\\
    &=H_0+H_1
\end{align}
where the term $H_0$ represents the linear contributions to the energy of the system, while $H_1$ is the first-order nonlinear contribution. \\
This Hamiltonian can be converted to its quantum form promoting the field $\Phi(x,t)$ to the quantum operator $\hat{\Phi}(x,t)$. \\
In direct analogy with (Eq. (\ref{eq:ClassicalPhases})) one can express the flux operator in terms of continuous-mode functions \cite{Vool2017}, such as $\hat{H}_0$ is diagonal in the plane-waves unperturbed modes decomposition: 
\begin{equation}
\label{eq:Continuous_Flux_Operator}
    \hat{\Phi}(x,t)=\displaystyle\sum_{\nu=L,R}\displaystyle\int^{\infty}_0\Bigg[\sqrt{\frac{\hbar l(x)}{4\pi k_{\omega}(x)}}\hat{a}_{\nu\omega}e^{\;i(\pm k_\omega(x) x-\omega t)}+\text{H.c.}\Bigg]d\omega
\end{equation}
where the subscript $R$ denotes a progressive wave, while $L$ denotes a regressive wave (i.e. $\hat{a}_{R\omega}$ represents the annihilation operator of a right-moving field of frequency $\omega$).
In Ref. \cite{Santos1995} it is demonstrated that replacing the definition given in (Eq. (\ref{eq:Continuous_Flux_Operator})) into the linear Hamiltonian, $\hat{H}_0$ takes the form
\begin{equation}
    \hat{H}_0=\displaystyle\sum_{\nu=R,L}\displaystyle\int^{\infty}_0\big[\hbar\omega\hat{a}^{\dag}_{\nu\omega}\hat{a}_{\nu\omega}\big]d\omega
\end{equation}
(where the zero-point energy, which doesn't influence the dynamics of the amplifier, has been omitted).\\
Using the expansion of $\hat{\Phi}(x,t)$ introduced above, under the hypothesis of a strong right moving classical pump centered in $\omega_p$, and that the fields $\hat{a}_{\nu\omega}$ are small except for frequencies closed to the pump frequency (i.e. replacing $\hat{a}_{\nu\omega}$ with $\hat{a}_{\nu\omega}+b(\omega)$, where $b(\omega)$ is a complex valued function centered in $\omega_p$), the nonlinear Hamiltonian $\hat{H}_1$ can be expressed under strong pump approximation, at the first-order in $b(\omega)$, as the sum of three different contributions: 
\begin{equation}
    \hat{H}_1=\hat{H}_{CPM}+\hat{H}_{SQ}+H_{SPM}
\end{equation}
In the expressions of these contribution the fast rotating terms and the highly phase mismatched left-moving field have been neglected:
\begin{equation}
    \hat{H}_{CPM}=-\frac{\hbar}{2\pi}\displaystyle\int^{\infty}_{0}d\omega_s d\omega_{i}d\Omega_p d\Omega_{p'}\sqrt{k_{\omega_s}k_{\omega_i}}\beta^*(\Omega_p)\beta(\Omega_{p'})\Upsilon(\omega_s,\omega_i,\Omega_p,\Omega_{p'})\hat{a}^{\dag}_{R\omega_s}\hat{a}_{R\omega_i}+\text{H.c}
\end{equation}
describes the cross-phase modulation,
\begin{equation}
    \hat{H}_{SQ}=-\frac{\hbar}{4\pi}\displaystyle\int^{\infty}_0 d\omega_s d\omega_i d\Omega_p d\Omega_{p'} \sqrt{k_{\omega_s}k_{\omega_i}}\beta(\Omega_p)\beta(\Omega_{p'})\Upsilon(\omega_s,\Omega_p,\omega_i,\Omega_{p'})\hat{a}^{\dag}_{R\omega_s}\hat{a}^{\dag}_{R\omega_i}+\text{H.c.}
\end{equation}
describes the broadband squeezing and
\begin{equation}
    H_{SPM}=-\frac{\hbar}{4\pi}\displaystyle\int^{\infty}_0 d\omega_s d\omega_i d\Omega_p d\Omega_{p'} \sqrt{k_{\omega_s}k{\omega_i}}\beta^*(\Omega_p)\beta(\Omega_{p'})\Upsilon(\omega_s,\omega_i,\Omega_p,\Omega_{p'}) b^{*}(\omega_s)b(\omega_{i})+\text{H.c.}
\end{equation}
describes the self-phase modulation. $\beta(\Omega)$ is the dimensionless pump amplitude, proportional to the ratio between the pump current $I_J(\Omega_p)$ and the critical current of the junctions $I_c$:
\begin{equation}
    \beta(\Omega_p)=\frac{I_J(\Omega_p)}{4I_c}
\end{equation}
The function $\Upsilon(\omega_1,\omega_2,\omega_3,\omega_4)$ is the phase matching function, defined as
\begin{equation}
    \Upsilon(\omega_1,\omega_2,\omega_3,\omega_4)=\displaystyle\int_{0}^{z} e^{-i(k_{\omega_1}(x)-k_{\omega_2}(x)+k_{\omega_3}(x)-k_{\omega_4}(x))}dx
\end{equation}
Assuming the nonlinear Hamiltonian $\hat{H}_1$ as a perturbative term of the Hamiltonian $\hat{H}_0$ for which the continuous modes are non-interacting, and assuming the initial time of the interaction $t_0=-\infty$ and the final time $t_1=+\infty$, it is possible to relate the input field of the system to the output one introducing the asymptotic output field
\begin{equation}
\label{eq:Asymptotic_Output_Field}
\hat{a}^{out}_{R\omega}=\hat{U}\hat{a}_{R\omega}\hat{U}^{\dag},
\end{equation}
where $\hat{U}$ is the asymptotic unitary evolution operator (approximated to the first order in $\hat{H}_1$) 
\begin{equation}
    \hat{U}\equiv\hat{U}(-\infty,\infty)=e^{-\frac{i}{\hbar}\hat{K}_1} \hspace{0.5cm}\text{where}\hspace{0.5cm}\hat{K}_1=\hat{K}_{CPM}+\hat{K}_{SQ}+K_{SPM}
\end{equation}
Working in the monochromatic degenerate pump limit $b(\Omega_{p'})=b(\Omega_p)\to b(\Omega_p)\delta(\Omega_p-\omega_p)\equiv b_p$ (where $\omega_p$ is the pump frequency) the propagators take the form
\begin{equation}
    \hat{K}_{CPM}=-2\hbar|\beta_p|^2 z\displaystyle\int^{\infty}_0  k_{\omega_s}\hat{a}^{\dag}_{R\omega_s}\hat{a}_{R\omega_s} d\omega_s
\end{equation}
\begin{equation}
    \hat{K}_{SQ}=-\frac{i\hbar}{2}|\beta_p|^2\displaystyle\int^{\infty}_{0}\sqrt{k_{\omega_s}k_{\omega_i}}\bigg[\frac{1}{\Delta k}\bigg(e^{i\Delta k z}-1\bigg)\Bigg]\hat{a}^{\dag}_{R\omega_s}\hat{a}^{\dag}_{R\omega_i}d\omega_s+\text{H.c.}
\end{equation}
and
\begin{equation}
    K_{SPM}=-\hbar z |\beta_p|^2 k_{\omega_p}b_p^{*}b_p
\end{equation}
where $\beta(\omega_p)\equiv\beta_p$ and $\Delta k=2k_{\omega_p}-k_{\omega_s}-k_{\omega_i}$ is the chromatic dispersion. \\
Similarly to the previous classical treatment, one can introduce the co-rotating framework by replacing the fields operators with
\begin{equation} 
\hat{a}_{R\omega_s}(z)=\hat{\Tilde{a}}_{R\omega_s}(z)\;e^{2i|\beta_p|^2k_{\omega_s} z}\hspace{0.5cm}\text{and}\hspace{0.5cm} b_p(z)=\Tilde{b}_p(z)\;e^{i|\beta_p|^2k_{\omega_p}z}
\end{equation}
In this framework, one can derive the following differential equation
\begin{equation}
\label{eq:CME_Q}
    \frac{\partial \hat{\Tilde{a}}_{R\omega_s}(z)}{\partial z}=\frac{i}{\hbar}\bigg[\frac{d}{dz}\hat{K}_1,\hat{\Tilde{a}}_{R\omega_s}\bigg]=i\beta_p^2\sqrt{k_{\omega_s}k_{\omega_i}}e^{i\Psi_2(\omega_s)z}\hat{\Tilde{a}}^{\dag}_{R\omega_i}
\end{equation}
and
\begin{equation}
\frac{\partial\Tilde{b}_p(z)}{\partial z}=-\bigg\{\frac{d K_{SPM}}{dz},\Tilde{b}_p\bigg\}=0
\end{equation}
where $\Psi_2 (\omega_s)=\Delta k+2|\beta_p|^2(k_{\omega_p}-k_{\omega_s}-k_{\omega_i})$ is the total phase mismatch. These latter are formally identical to (Eq. (\ref{eq:CME_1})) and (Eq. (\ref{eq:CME_2})), up to a frequency-dependent normalization of the wave-amplitudes, under the undepleted pump approximation.
Ref. \cite{O'Brien2014} derives an exact solution for (Eq. (\ref{eq:CME_Q})), being 
\begin{align}
\label{eq:Final_2}
    \hat{\Tilde{a}}^{out}_{R\omega_s}(z)=\bigg[\bigg(\cosh\Big({g_2(\omega_s)z}\Big)-&\frac{i\Psi_2(\omega_s)}{2g_2(\omega_s)}\sinh{\Big(g_2(\omega_s)z\Big)}\bigg)\hat{\Tilde{a}}_{R\omega_s}+\nonumber\\
    &+i \frac{\beta^2_p\sqrt{k_{\omega_s}k_{\omega_i}}}{g_2(\omega_s)}\sinh{\Big(g_2(\omega_s)z\Big)}\;\hat{\Tilde{a}}^{\dag}_{R\omega_i}\bigg]e^{i\frac{\Psi_2(\omega_s)}{2}z}
\end{align}
where
\begin{equation}
    g_2(\omega_s)=\sqrt{|\beta^2_p|^2k_{\omega_s}k_{\omega_i}-\bigg(\frac{\Psi_2(\omega_s)}{2}\bigg)^2}
\end{equation}
If a state moves inside a TWJPA of length $z=aN$, the power gain will be $G(\omega_s,aN)=\big<\hat{\Tilde{a}}^{out}_{R\omega_s}(aN)\big|\hat{\Tilde{a}}^{out^\dag}_{R\omega_s}(aN)\big>/\big<\hat{\Tilde{a}}_{R\omega_s}\big|\hat{\Tilde{a}}^{\dag}_{R\omega_s}\big>$.


\subsection{Quantum Hamiltonian model based on discrete-mode operators}
An alternative approach for the derivation of the quantum dynamics of a TWJPA is the one proposed in Ref.~\cite{vanderReep2019}. In this model the quantum Hamiltonian for a 4WM parametric amplifier is expressed as the integral, along an arbitrary quantization length $l_q$, of the linear energy density stored in each element of the circuit. The energy stored per unit length in a Josephson junction can be derived from (Eq. (\ref{eq:Josephson_Energy})) dividing each term by the elementary cell length $a$ and replacing $\Delta\Phi_n$ with its continuous counterpart. Instead, the energy stored per unit length in a capacitance C can be alternatively expressed in terms of flux difference $\Delta\Phi$ or stored charge Q as
\begin{align}
    U_C=\frac{1}{a}\displaystyle\int^t_{t_0} VI\;dt'
    &=\frac{1}{a}\displaystyle\int^t_{t_0}\frac{d\Delta\Phi}{dt'}\;C\frac{d}{dt'}\bigg[\frac{d\Delta\Phi}{dt'}\bigg]\;dt'=\frac{1}{2}\frac{C}{a}\Delta\Phi(t)\frac{\partial^2\Delta\Phi}{\partial t^2}\\
    &=\frac{1}{a}\displaystyle\int^t_{t_0}\frac{Q}{C}\frac{dQ}{dt'}\;dt'=\frac{1}{2a}\frac{1}{C}Q^2(t)
\end{align}
under the assumption that $\Delta\Phi(t_0)=0$ and $Q(t_0)=0$.\\
Therefore, the quantum Hamiltonian of the system can be expressed, with an approximation to the first non-linear order, as 
\begin{align}
\label{eq:Hamiltonian}
    \hat{H}&=\displaystyle\int_{l_q} \Big[U_J+U_{C_J}+U_{C_0}\Big]dx\approx\nonumber\\
    &\approx\displaystyle\int_{l_q}\bigg[\bigg(\frac{1}{2aL_{J_0}}\Delta\hat{\Phi}-\frac{1}{24aL_{J_0}}\Big(\frac{2\pi}{\Phi_0}\Big)^2\Delta\hat{\Phi}^3+\frac{1}{2}\frac{C_J}{a}\frac{\partial^2\Delta\hat{\Phi}}{\partial t^2}\bigg)\Delta\hat{\Phi}+\frac{1}{2a}\frac{1}{C_0}\hat{Q}^2_{C_0}\bigg]\;dx
\end{align}
where $\hat{Q}$ and $\hat{\Phi}$ are quantum operators. The former can be expressed, as suggested in \cite{Vool2017} and adapted for discrete-mode operators in \cite{Loudon2000}, as
\begin{equation}
\label{eq:Charge_Operator}
    \hat{Q}_{C_0}=\displaystyle\sum_{n}\frac{C_0}{a}\hat{V}_{C_0,n}=\displaystyle\sum_{n}\frac{C_0}{a}\sqrt{\frac{\hbar\omega_n a}{2C_0l_q}}\;(\hat{a}_n\;e^{i(k_nx-\omega_n t)}+\text{H.c.})
\end{equation}
(here $k_{-n}=-k_n$ and $\omega_{-n}=\omega_n$).\\
Before defining the flux operator, it is necessary to define an effective inductance $L_{eff}$ of the transmission line (modeled, as shown in Fig. \ref{fig:Elementary_cell}, as a parallel of the nonlinear Josephson inductance $L_J$ and the capacitance $C_J$):
\begin{equation}
    \frac{1}{j\omega_n L_{eff}}=\frac{1}{j\omega_n L_{J}}+j\omega_n C_J \hspace{0.5cm}\text{hence}\hspace{0.5cm}L_{eff}=\frac{L_J}{1-\omega^2L_J C_J}\equiv L_J\Lambda_n
\end{equation}
Using the telegrapher's equation \cite{Pozar2012} the discrete-mode current operator, under slowly varying amplitude approximation ($\partial\hat{a}_n/\partial x\approx 0$), can be derived from the voltage one as:
\begin{equation}
    \hat{I}_{L_{eff}}=\displaystyle\sum_{n} \text{sgn}(n)\sqrt{\frac{\hbar\omega_n a}{2L_{eff} l_q}}\;(\hat{a}_n\;e^{i(k_nx-\omega_n t)}+\text{H.c.})
\end{equation}
Therefore, the flux operator can be expressed as
\begin{equation}
\label{eq:Flux_Operator_2}
    \Delta\hat{\Phi}=\frac{L_J}{a}\hat{I}_{L_{eff}}\hspace{0.5cm} \text{where} \hspace{0.5cm}L_J(\Delta\Phi)=\frac{\Delta\Phi}{I_J}=\frac{\frac{2\pi}{\Phi_0}\Delta\Phi}{I_c\frac{2\pi}{\Phi_0}\sin{\Big(\frac{2\pi}{\Phi_0}\Delta\Phi\Big)}}=L_{J_0}\frac{\frac{2\pi}{\Phi_0}\Delta\Phi}{\sin{\Big(\frac{2\pi}{\Phi_0}\Delta\Phi\Big)}}
\end{equation}
The recursive relation deriving from (Eq. (\ref{eq:Flux_Operator_2})) can be solved iteratively. Exploiting a power series expansion of the sine function and considering just the first order of interaction, it results that
\begin{equation}
\label{eq:Flux_Operator}
    \Delta\hat{\Phi}=\displaystyle\sum_n\Bigg[1+\frac{\Lambda_n}{12}\Bigg(\frac{2\pi}{\Phi_0}\Delta\hat{\Phi}^{(0)}\Bigg)^2+O\Bigg[\Bigg(\frac{2\pi}{\Phi_0}\Delta\hat{\Phi}^{(0)}\Bigg)^4\Bigg]\Bigg]\Delta\hat{\Phi}^{(0)}_n
\end{equation}
where $\Delta\hat{\Phi}^{(0)}$ is the zero-order approximation of the flux quantum operator
\begin{equation}
    \Delta\hat{\Phi}^{(0)}=\displaystyle\sum_{n}\Delta\hat{\Phi}^{(0)}_n=\displaystyle\sum_n\frac{k_n a}{\omega_n}\sqrt{\frac{\hbar\omega_n a}{2C_0l_q}}\;(\hat{a}_n\;e^{i(k_n x-\omega_n t)}+\text{H.c.})
\end{equation}
Substituting (Eq. (\ref{eq:Charge_Operator})) and (Eq. (\ref{eq:Flux_Operator})) in (Eq. (\ref{eq:Hamiltonian})), and limiting the expression to the first non linear order, the Hamiltonian for a 4WM amplifiers turns up
\begin{align}
    \hat{H}&=\displaystyle\sum_{n}\hbar\omega_n\bigg(\hat{a}^{\dag}_n\hat{a}_n+\frac{1}{2}\bigg)+\displaystyle\sum_{n,m,l,k}\frac{-i\hbar^2a}{96L_{J_0}I^2_cl^2_q\Delta k_{nmlk}}\;e^{-i\Delta\omega_{nmlk}t}(e^{i\Delta k_{nmlk} l_q}-1)\times\nonumber\\
    &\times \Big\{(1-4L_{J_0}\Lambda_n C_J\omega^2_k)\big(\hat{\Tilde{a}}+\text{H.c.}\big)_{n\times m\times l\times k}+4L_{J_0}\Lambda_n C_J\big[2\big(\omega(-i\hat{\Tilde{a}}+\text{H.c.})\big)_{n\times m}\times\nonumber\\
    \label{eq:Hamiltonian_2}
    &\times\big(\hat{\Tilde{a}}+\text{H.c.}\big)_{l\times k}+\big(\hat{\Tilde{a}}+\text{H.c}\big)_{n\times m}\big(\omega(-i\hat{\Tilde{a}}+\text{H.c.})\big)_{l\times k}\big]\Big\}
\end{align}
where $\hat{\Tilde{a}}\equiv \text{sgn}(n)\sqrt{\Lambda_n\omega_n}\;\hat{a}_n$, $\Delta k_{nmlk}\equiv\pm k_n\pm k_m \pm k_l \pm k_k$, $\Delta\omega_{nmlk}\equiv\pm\omega_n\pm\omega_m\pm\omega_l\pm\omega_k$ (a plus (minus) sign refers to a corresponding annihilation (creation) operator) and the subscript $i\times j$ indicates a multiplication (i.e.$\big(\Lambda\omega\big)_{i\times j}= \Lambda_i\omega_i\Lambda_j\omega_j$).\\
Neglecting the constant zero-point energy and assuming a strong degenerate classical pump (as shown in Ref. \cite{Loudon2000}) 
\begin{equation}
    \hat{a}_p\to-i\sqrt{\frac{\omega_p C_0 l_q}{2\hbar a}}A_p
\end{equation}
it is possible to approximate the Hamiltonian in (Eq. (\ref{eq:Hamiltonian_2})) to the second order in $\hat{a}^{(\dag)}_{s,i}$ as
\begin{equation}
\label{eq:Hamiltonian_vdR_CP}
    \hat{H}^{(CP)}\approx\displaystyle\sum_{n=s,i}\hbar\big(\omega_n+\xi_n|A_p|^2\big)\hat{a}^{\dag}_n\hat{a}_n-\hbar\big(\chi A^2_p\hat{a}^{\dag}_s\hat{a}^{\dag}_i+\text{H.c.}\big)
\end{equation}
where
\begin{equation}
    \xi_n=\frac{k_p^2a^2\Lambda_n\omega_n}{32I^2_cL^2_{J_0}}(4-3\delta_{pn})\Bigg(1+\frac{2}{3}\Bigg(\frac{\Lambda_p}{\Lambda_n}+\frac{\Lambda_n}{\Lambda_p}-2\Bigg)\Bigg)
\end{equation}
represents the quantum self-phase modulation (when $n=p$) and the quantum cross-phase modulation (when $n=s,i$), whereas the coupling constant $\chi$ is defined as
\begin{align}
    \chi=\frac{k^2_pa^2\sqrt{\Lambda_s\omega_s\Lambda_i\omega_i}}{16I^2_c L^2_{J_0}}&\Bigg(1+\frac{L_{J_0} C_J}{6}\Big[\omega_p\omega_s(-2\Lambda_p+5\Lambda_s-3\Lambda_i)+\nonumber\\
    &+\omega_p\omega_i(-2\Lambda_p-3\Lambda_s+5\Lambda_i)+\omega_s\omega_i(4\Lambda_p-2\Lambda_s-2\Lambda_i)\Big]\Bigg)
\end{align}
Starting from the Hamiltonian $\hat{H}^{(CP)}$ it is possible to calculate the Heisenberg equation of motion for the classical pump amplitude and for the quantum operators $\hat{a}_s$ and $\hat{a}_i$, obtaining the coupled mode equations:
\begin{equation}
\label{eq:CME_vdR_1}
    \frac{\partial A_p}{\partial t}=-i\big(\omega_p+2\xi_p|A_{p}|^2\big)A_p+2i\chi^{*}A^{*}_p\hat{a}_s\hat{a}_i
\end{equation}
\begin{equation}
\label{eq:CME_vdR_2}
    \frac{\partial\hat{a}_{s(i)}}{\partial t}=-i(\omega_{s(i)}+\xi_{s(i)}|A_p|^2)\hat{a}_{s(i)}+i\chi A^2_p\hat{a}^{\dag}_{i(s)}
\end{equation}
In Ref. \cite{vanderReep2019} the hypothesis under which the classical coupled mode equations (Eq. (\ref{eq:CME_1})) and (\ref{eq:CME_2}) can be obtained from (Eq. (\ref{eq:CME_vdR_1})) and (Eq. (\ref{eq:CME_vdR_2})) is described in detail.\\
Moving to a co-rotating frame ($\hat{a}_{s(i)}\to\hat{a}_{s(i)}e^{i\xi_{s(i)}|A_{p_0}|^2 z}$) the Hamiltonian (Eq. (\ref{eq:Hamiltonian_vdR_CP})) can be expressed as
\begin{equation}
    \hat{H}^{CP}_{rot}=-\hbar\Big(\chi|A_p|^2\hat{a}^{\dag}_s\hat{a}^{\dag}_ie^{-i\Psi^{'}_3 t}+\text{H.c.}\Big)
\end{equation}
where $\Psi^{'}_3=(4\xi_p-\xi_s-\xi_i) |A_{p_0}|^2$. In this frame, introducing the undepleted pump assumption, (Eq. (\ref{eq:CME_vdR_2})) turns into
\begin{equation}
    \frac{\partial\hat{a}_{s(i)}}{\partial t}=i\chi|A_{p_0}|^2\hat{a}^{\dag}_{i(s)}e^{-i\Psi^{'}_3 t}
\end{equation}
whose solutions are
\begin{equation}
\label{eq:Final_3}
    \hat{a}_{s(i)}(t)=\bigg[\hat{a}_{s(i)_0}\bigg(\cosh{\big(g_3^{'}t\big)}+\frac{i\Psi^{'}_3}{2g^{'}_3}\sinh{\big(g^{'}_3t\big)}\bigg)+\frac{i\chi|A_{p_0}|^2}{g^{'}_3}\hat{a}^{\dag}_{i(s)_0}\sinh{\big(g^{'}_3 t\big)}\bigg]
\end{equation}
where the exponential complex gain factor is defined as 
\begin{equation}
    g^{'}_3=\sqrt{|\chi|^2|A_{p_0}|^4-\bigg(\frac{\Psi^{'}_3}{2}\bigg)^2}
\end{equation}
If a state spends a time $t$ in the amplifier, the gain can be expressed as $G^Q_s(t)=\big<\hat{a}^{\dag}_s(t)\hat{a}_s(t)\big>/\big<\hat{a}^{\dag}_{s_0}\hat{a}_{s_0}\big>$.\\
To make the results of this last treatment, in which the operators are expressed as a function of the time, comparable with the previous ones, in which the operators are expressed as a function of the space coordinate, we need to take into account the phase velocity of the tones. It turns out that:
\begin{equation}
    \Psi_3=\Delta k +\bigg(4\xi_p\frac{\omega_p}{|k_p|}-\xi_s\frac{\omega_s}{|k_s|}-\xi_i\frac{\omega_i}{|k_i|}\bigg)\hspace{0.5cm}\text{and}\hspace{0.5cm}g_3=\sqrt{|\chi|^2|A_{p_0}|^4\bigg(\frac{\omega_p}{|k_p|}\bigg)^2-\bigg(\frac{\Psi_3}{2}\bigg)^2}
\end{equation}
where $\Delta k=2k_p-k_s-k_i$ is the chromatic dispersion.

\section{Conclusions}
\begin{figure}[h!]
 \begin{centering}
    \includegraphics[width=1\textwidth]{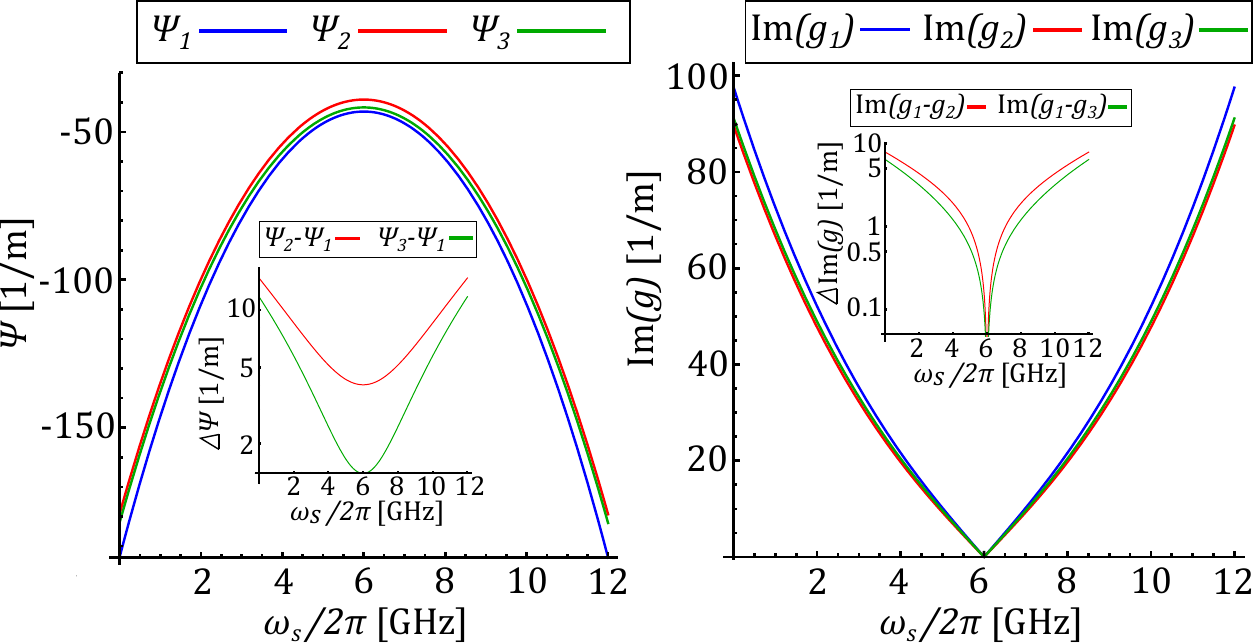}
\par\end{centering}
    \caption{Comparison of total phase mismatches ($\Psi_i$) and exponential complex gain factors ($g_i$) predicted by the three treatments presented in section \ref{sec:models}. For these calculations typical constructive parameters have been assumed: $a=50\;\mu$m, $I_c=5\;\mu$A, $C_J=300$ fF and $C_0=35$ fF, in such a way that the characteristic impedance is $Z\approx50\;\Omega$. Furthermore $\omega_p/2\pi=6$ GHz and $I_p=Ic/2$. In the insets, the differences between the quantum predictions and the classical ones are presented.}
    \label{fig:Psi&g}
\end{figure}

In the present chapter we have presented the state-of-the-art of the experimental evidences in the field of Josephson junctions-based travelling-wave metamaterials through an historical review in section \ref{sec:history}. Moreover, in section \ref{sec:models}, we have reported three different theoretical approaches for the prediction of a TJWPA dynamics, in the particular case of a 4WM process. Assuming similar simplifying hypothesis, like the use of a classical undepleted degenerate pump, the presence of slowly varying fields along the transmission line and approximating the nonlinearities of the system up to the first order, a similar expression for the signal amplitude (or field annihilation/creation operators in the case of quantum theories) expressed in a co-rotating frame, is derived in the three treatments. The three exponential complex gain factors ($g_i$) and the three total phase mismatches ($\Psi_i$) derived in these models are analytically different but numerically similar, as shown in Figure \ref{fig:Psi&g} (where the two insets report the differences between the quantum predictions and the classical ones).
\\
It is important to observe that in (Eq. (\ref{eq:Final_1})), (Eq. (\ref{eq:Final_2})) and (Eq. (\ref{eq:Final_3})) the last term is always equal to zero in the case of a zero initial idler amplitude. In such a case, and under the hypothesis of a perfect phase matching ($\Psi_i=0$), $g_i$ is real and the amplification gain increases exponentially with the line length whereas, in the case of a non-zero phase mismatch, $g_i$ is imaginary and the gain increases quadratically \cite{O'Brien2014}.\\
Although the results of the quantum theories are similar to the classical ones, the description of the system dynamics with a quantum theory grants the possibility to evaluate photon-number distributions, squeezing effects and avarages, standard deviations or higher-order moments of the measurements operators, taking into account the commutation relations between operators explicitly. For istance, detailed calculations of the output state of a TWJPA in the case of a single-photon input state and in the case of a coherent input state are presented in \cite{vanderReep2019}.\\

\section{Acknowledgments}
The author would like to thank Luca Callegaro for
the stimulating discussion. This work was partially funded by the Joint
Research Project PARAWAVE of the European Metrology Programme for Innovation and Research (EMPIR). This project has received funding from the EMPIR programme co-financed by the Participating States and from the European Unions Horizon 2020 research and innovation programme.

\end{document}